\documentclass{PoS}

\title{The nature of the finite temperature QCD transition as a function of the quark masses}

\ShortTitle{The nature of the finite temperature QCD transition as a function of the quark masses}

\author{\speaker{G.~Endr\H{o}di}$^1$, Z. Fodor$^{1,2}$, S.D.~Katz$^{1,2}$, K.K. Szab\'o$^2$\\
$^1$Institute for Theoretical Physics, E\"otv\"os University, H-1117 Budapest, Hungary. \\
$^2$Department of Physics, University of Wuppertal, D-42097 Wuppertal, Germany.\\
       E-mail: \email{endrodi@general.elte.hu}
}

\abstract{The finite temperature QCD transition for physical quark masses is a crossover. For smaller quark masses a first-order phase transition is expected. Using Symanzik improved gauge and stout improved fermion action for 2+1 flavour staggered QCD we give estimates/bounds for the phase line separating the first-order region from the crossover one. The calculations are carried out on two different lattice spacings. Our conclusion for the critical mass is $m_0 \lesssim 0.07 \cdot m_{phys}$ for $N_T=4$ and $m_0 \lesssim 0.12 \cdot m_{phys}$ for $N_T=6$ lattices.}

\FullConference{The XXV International Symposium on Lattice Field Theory\\
                July 30 - August 4 2007\\
                Regensburg, Germany}

\begin{document}

\section{Introduction}

Quantum chromodynamics (QCD) is the theory of strong interactions. Due to one of its most important properties, asymptotic freedom, at high temperatures it describes a different phase of matter called quark-gluon plasma (QGP). The phase transition between the hadronic phase of matter and QGP can be investigated by lattice simulations. The transition at zero chemical potential -- which represents the case of equal number of quarks and antiquarks -- is of huge importance, since it is relevant for both regarding the early Universe and high energy collisions.

\begin{figure}[h]
\centering
\includegraphics[height=5cm,width=5cm]{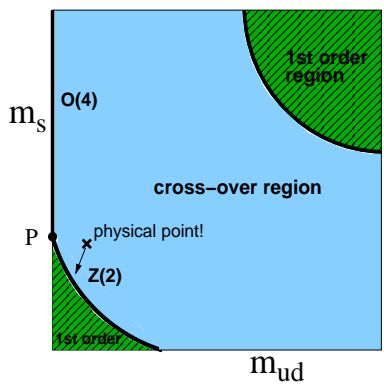}
\caption{The phase diagram of QCD. First-order and crossover regions are separated by second-order lines. For the one pointed towards by the arrow we expect a Z(2) universality class. The exact position of the line is to be determined.}
\label{fig:quarkmass}
\end{figure}

The 2+1 flavour QCD transition was recently found to be an analytic crossover~\cite{Aoki:2006} (instead of a first-order phase transition), which usually results in different transition temperatures for different observables~\cite{Aoki:2006br} and to a broadening of the equation of state around the transition temperature~\cite{Aoki:2005vt}. These works were carried out using physical quark masses; nevertheless different values of the quark masses can also have relevance. For three massless quarks we expect from QCD effective models, that a first-order phase transition takes place. For infinite quark masses (which describe pure gauge theory) lattice results indicate that there is also a first-order transition. For two massless flavours a second-order transition is expected. We can summarize our knowledge on figure \ref{fig:quarkmass}. 

There are second-order phase transition lines, that separate the first-order and crossover regions. For the 2 flavour case, the universality class of the phase line is predicted to be O(4)~\cite{Pisarski:1983ms}, while for the 2+1 case, we expect Z(2). However, the exact position of this latter phase line still needs to be determined with adequate accuracy. In the work of~\cite{deForcrand:2006pv} the phase line is found to be at about $80\%$ of the physical quark mass on $N_T=4$ lattices with the unimproved staggered action. The same authors have presented their $N_T=6$ study at the present conference~\cite{ForPhil:lat07}. Based on the results about the strength of the transition for different lattice spacings and/or discretization schemes, one expects that reducing the discretization errors results in a weaker transition. In agreement with this expectation they observed that the first-order region shrinks, and the physical point is farther from the phase line.

The location of the second-order line has high importance, since combined with the curvature of the phase surface (in the $m-\mu$ space), it can influence the position~\cite{Fodor:2001pe, Fodor:2004nz} (or even the existence~\cite{deForcrand:2006pv}) of the critical endpoint on the QCD phase diagram.

In this paper we give estimates and upper bounds for the critical mass by means of analyzing the behaviour of some quantities that are sensitive to the nature of the phase transition. These quantities will be the susceptibility and the Binder cumulant of the chiral condensate. It will turn out, that the latter is more trustworthy in locating the second-order line, nevertheless, we present results here regarding both quantities.

\section{Second-order behaviour} 

First we discuss what kind of behaviour we expect from the susceptibility and from the Binder cumulant in the vicinity of a second-order line. Then we carry out lattice simulations for different quark masses, and compare them with the expectations.

\subsection{The chiral susceptibility}

The susceptibility of the chiral condensate is defined as $\chi_{\bar\psi \psi} \equiv {\partial \bar\psi \psi \over \partial m}$. At the transition temperature it is supposed to show a pronounced peak. Distinguishing between first-order, second-order transitions and crossovers can be achieved by finite-size scaling of some properties of this peak. Particularly, for second-order transitions, the height of the peak should diverge at the critical point. We can observe this behaviour in a statistical physical approach. Our order parameter of the transition is the chiral condensate $\bar\psi \psi$, the reduced temperature is $t \equiv (T-T_C)/T_C$, and the external field, which breaks the symmetry is the quark mass $m$. The definitions of the critical indices $\epsilon, \gamma$, and $\delta$ are:
\begin{equation}
\bar\psi \psi \sim |t|^{\epsilon}, \quad \chi_{\bar\psi \psi} \sim |t|^{-\gamma}, \quad \bar\psi \psi^{\delta} |_{t=0}\sim m
\label{eq:critind}
\end{equation}

Now let's take the derivative of the last proportionality with respect to $m$, so the susceptibility can be expressed as a function of the quark mass. This will determine how the height of the peak grows while reducing the mass.
\begin{equation}
\label{eq:index_h}
\chi_{\bar\psi \psi} |_{t=0} \sim m^{{1 \over \delta} -1}
\end{equation}
From the first and third proportionality in (\ref{eq:critind}) we can also obtain how the critical temperature depends on the quark mass:
\begin{equation}
\label{eq:index_p}
|t| \sim m^{1 \over {\epsilon \delta}}
\end{equation}
It is worth mentioning that if we start from the second proportionality in (\ref{eq:critind}) and from (\ref{eq:index_h}), then we obtain $|t| \sim m^{\delta -1 \over \gamma \delta}$, which is identical to (\ref{eq:index_p}) (c.f. the $\gamma=\epsilon(\delta - 1)$ scaling rule).

In the following we will analyze the susceptibility peak as a function of $\beta \equiv 6 / g^2$. Since we restrict ourselves to the interval around the critical temperature, where the function $\beta(T)$ can be linearized, this means that in the above formulae we can substitute the reduced temperature with (the reduced) $\beta$.

The critical exponents in question can be looked up in the literature for the interesting universality classes; these values are summarized in the next table~\cite{kanaya:1995}:

\begin{center}
\begin{tabular}{c|c|c}
\label{tab:critexp}
& $1/\epsilon\delta$ & $1/\delta -1$ \\
\hline
\hline
3D Ising & 0.633 & -0.785 \\
3D O(2) & 0.598 & -0.794 \\
3D O(4) & 0.537 & -0.794 \\
\end{tabular}
\end{center}

\subsection{The Binder cumulant}
\label{sec:binder}

The cumulant is another useful quantity to distinguish between different types of phase transitions. Roughly, it measures how much the distribution of the order parameter is of Gaussian type. Its definition for the chiral condensate is as follows:

\begin{equation}
B_{\bar\psi \psi} \equiv {\langle (\delta \bar\psi \psi)^4 \rangle \over \langle (\delta \bar\psi \psi)^2 \rangle ^2}
\end{equation}
where $\delta$ denotes the deviation from the average, so $\delta \bar\psi \psi \equiv \bar \psi \psi - \langle \bar\psi \psi \rangle$. The actual value of the cumulant can be easily calculated for different distributions. We should analyze the distribution of $\bar\psi \psi$ in the infinite volume limit, at the critical temperature. For a first-order transition the distribution consists of two Dirac-deltas, for which $B_{\bar\psi \psi}=1$. For a crossover we have one Dirac-delta, which is (through a series of finite volumes) approached by Gaussian functions getting narrower and narrower. In this case $B_{\bar\psi \psi} \approx 3$. For second-order transitions the value of the cumulant depends on the universality class: for Z(2), that of the three-dimensional Ising-model, $B_{\bar\psi \psi}=1.604$~\cite{blote:1995}; for O(2), $B_{\bar\psi \psi}=1.242$~\cite{holtmann:2002}; while for O(4), $B_{\bar\psi \psi}=1.092$~\cite{kanaya:1995}.

\section{Results}

Our results were obtained by lattice simulations with 2+1 flavours of staggered quarks. We used Symanzik improved gauge and stout improved fermionic action; the details concerning the action and the simulation techniques are described in~\cite{Aoki:2006, Aoki:2006br, Aoki:2005vt}. For smaller volumes (ranging from $10^3 \times 4$ to $16^3 \times 4$) up to 500-1000 configurations were generated. For larger volumes (up to $24^3 \times 4$ and $N_T=6$ simulations) we had smaller statistics, about a few hundred configurations. Autocorrelation time was measured to be around 5, so we used every fifth configuration for measurements. Measurement of the chiral condensate was carried out with 60 random vectors.

In order to approach the second-order line we had to carry out simulations at very small quark masses. There are, however, limitations that we have to take into account. First of all, smaller masses increase CPU time by a factor of $1/m$. Still very important is, that we have to keep the lattice sizes much larger than the characteristic length of the system. This length is given by the inverse of the pion mass: $\ell_{\bar \psi \psi} = 1/m_{\pi} \approx 1/\sqrt{m}$, so for smaller masses we also needed larger lattices. Paying attention to these phenomena, we carried out simulations for quark masses ranging from $200 \%$ down to $5 \%$ of their physical values. 

\subsection{The chiral susceptibility}

\begin{figure}[h]
\vspace*{-0.5cm}
\centering
\includegraphics[height=5.5cm,width=6.5cm]{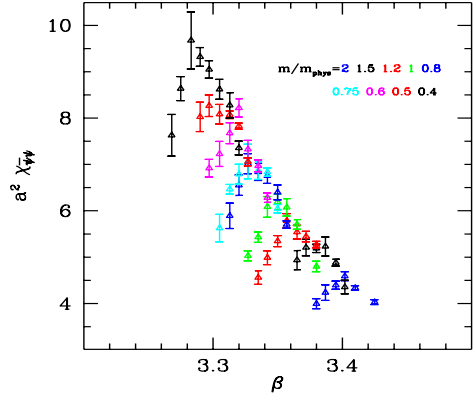}
\caption{The susceptibility peaks for different quark masses. Apparently the transition gets stronger for smaller masses, as it is shown by higher peaks.}
\label{fig:susc_mass}
\vspace*{-0.3cm}
\end{figure}

For every quark mass we used, we had to search for the susceptibility peak on the $\chi_{\bar\psi \psi} - \beta$ plane. These peaks are plotted on figure \ref{fig:susc_mass}., for the case of $m/m_{phys}=0.4 \ldots 2$. The height of the peak increases as smaller quark masses are used, which indicates the strengthening of the transition. 

\begin{figure}[h]
\centering
\mbox{
	\includegraphics*[height=5cm]{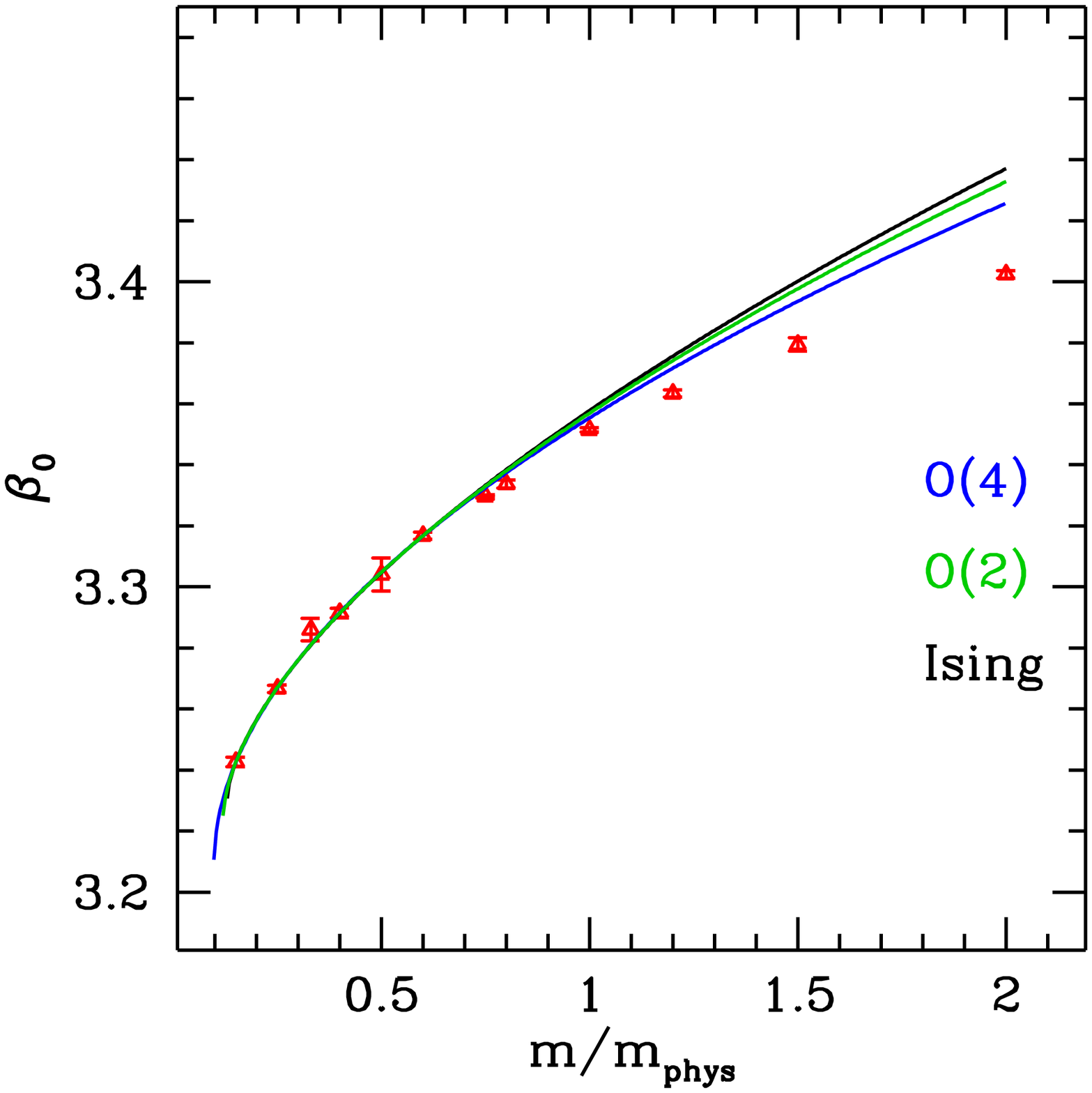}
     	\includegraphics*[height=5cm]{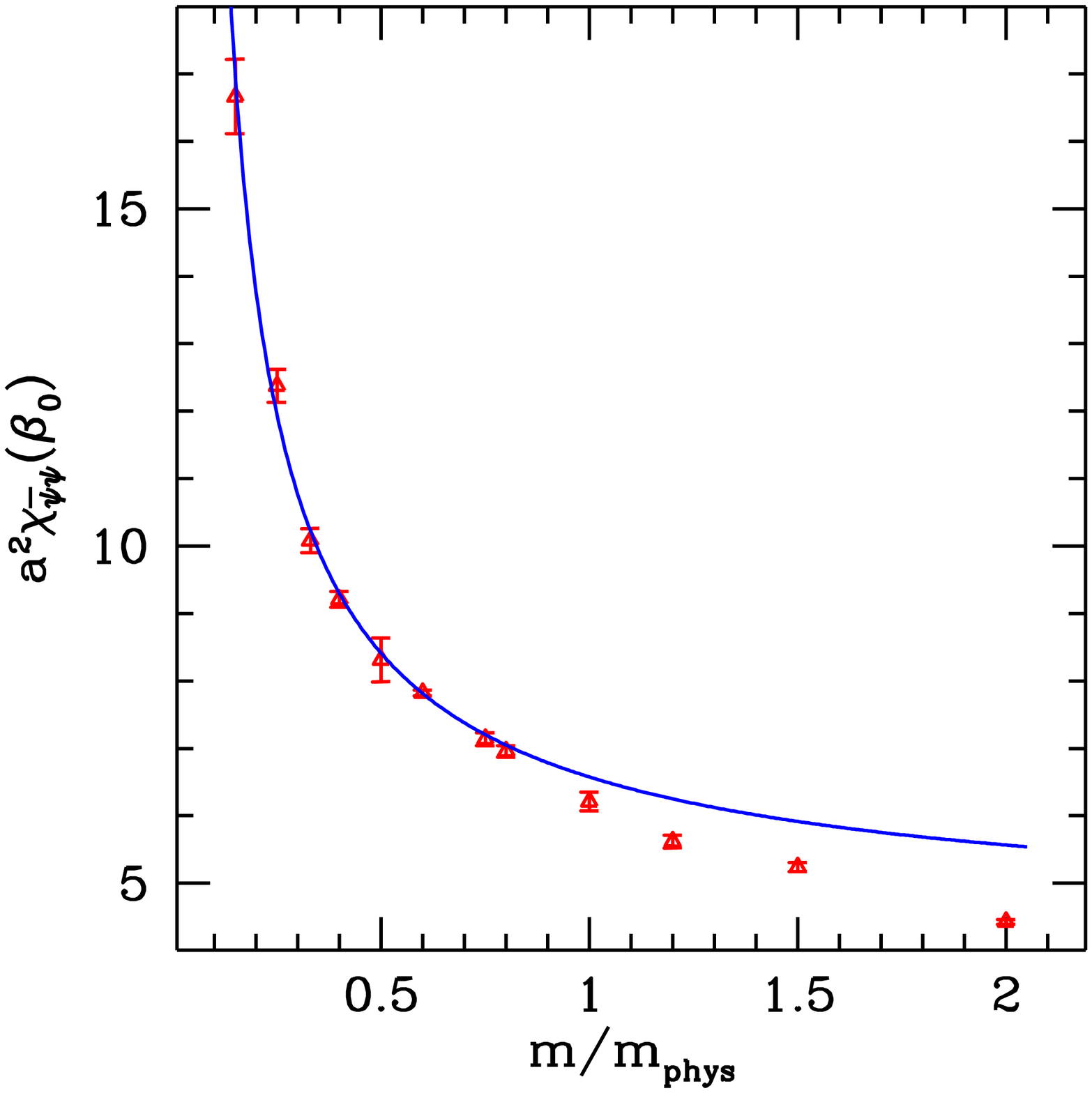}
     	\includegraphics*[height=5cm]{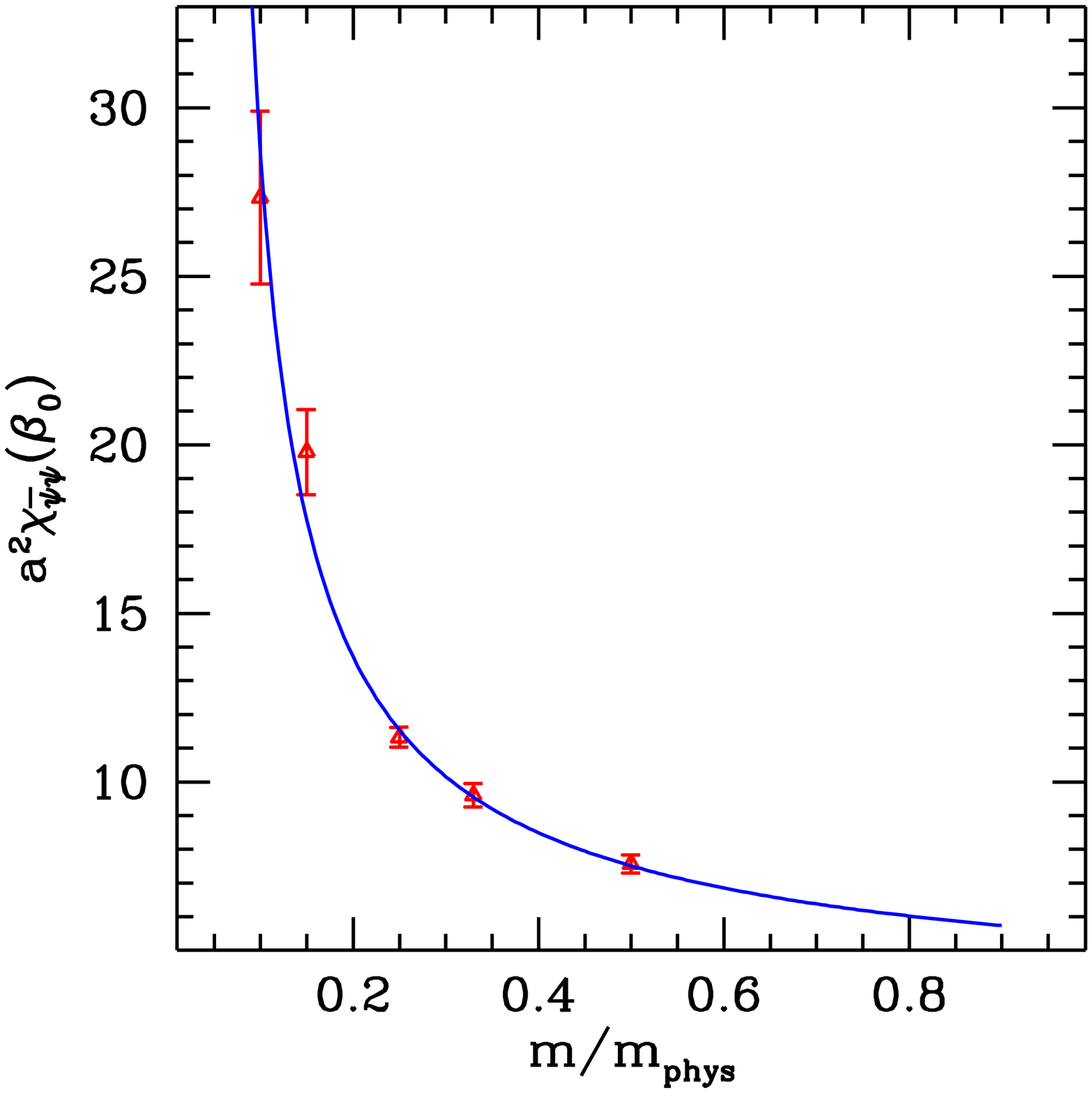}
}
\caption{The position (left panel) and the height (middle panel for $N_T=4$, right panel for $N_T=6$) of the susceptibility peak as a function of the quark mass. With solid lines shown are the power function fits, with exponents fixed at values from different universality classes (which give the same for the height).}
\label{fig:susc_props}
\end{figure}

As shown by equations (\ref{eq:index_h}) and (\ref{eq:index_p}), the height and the position of the susceptibility peak should follow a power-like behaviour, which has a singular point (non-analytical point for the case of the position) at the critical mass of the second-order point, denoted in the following by $m_0$. The critical indeces for these power functions (as summarized in table \ref{tab:critexp}.) are rather close to each other, particularly for the case of the height. This means that it is very difficult to distinguish between different universality classes from observing the behaviour of these quantities. However, if we suppose that that we are dealing with a given universality class (namely Z(2) for our case), than we may keep the exponent of the power function fixed, and perform a fit for the critical mass. These fits are shown on figure \ref{fig:susc_props}.

\begin{figure}[h]
\vspace*{-0.5cm}
\centering
\includegraphics[height=6cm,width=6.5cm]{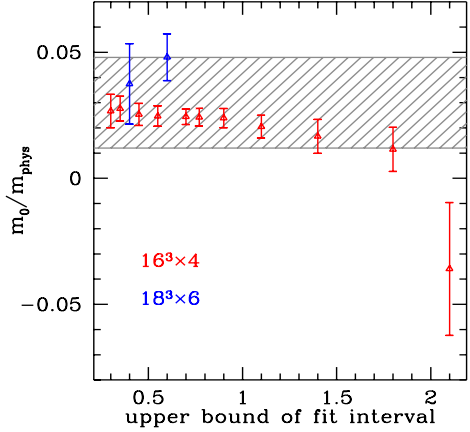}
\caption{The singular point of the power-function fitted to the height of the peak, versus the upper bound of the fit interval. As largest masses are excluded from the fit, we get deeper in the dominant region, where the behaviour of the susceptibility is governed by the appropriate critical index.}
\label{fig:fit_m0}
\end{figure}

We performed these fits for different fit intervals. The results from the height of the peak can be seen on figure \ref{fig:fit_m0}. As we narrow the fit interval by excluding points with largest masses, the estimate for $m_0$ reaches a nice plateu, which indicates that we are already in the dominant region of the second-order point at smaller masses. We can obtain an upper bound from this analysis, which is $m_0\lesssim 0.05 \cdot m_{phys}$. The same procedure was done also for the case of the position of the peak, for which the fits turned out to be less stable. Nevertheless from this latter we had $m_0\lesssim 0.12 \cdot m_{phys}$.

\subsection{The Binder cumulant}

\begin{figure}[h]
\centering
\mbox{
	\includegraphics*[height=5cm]{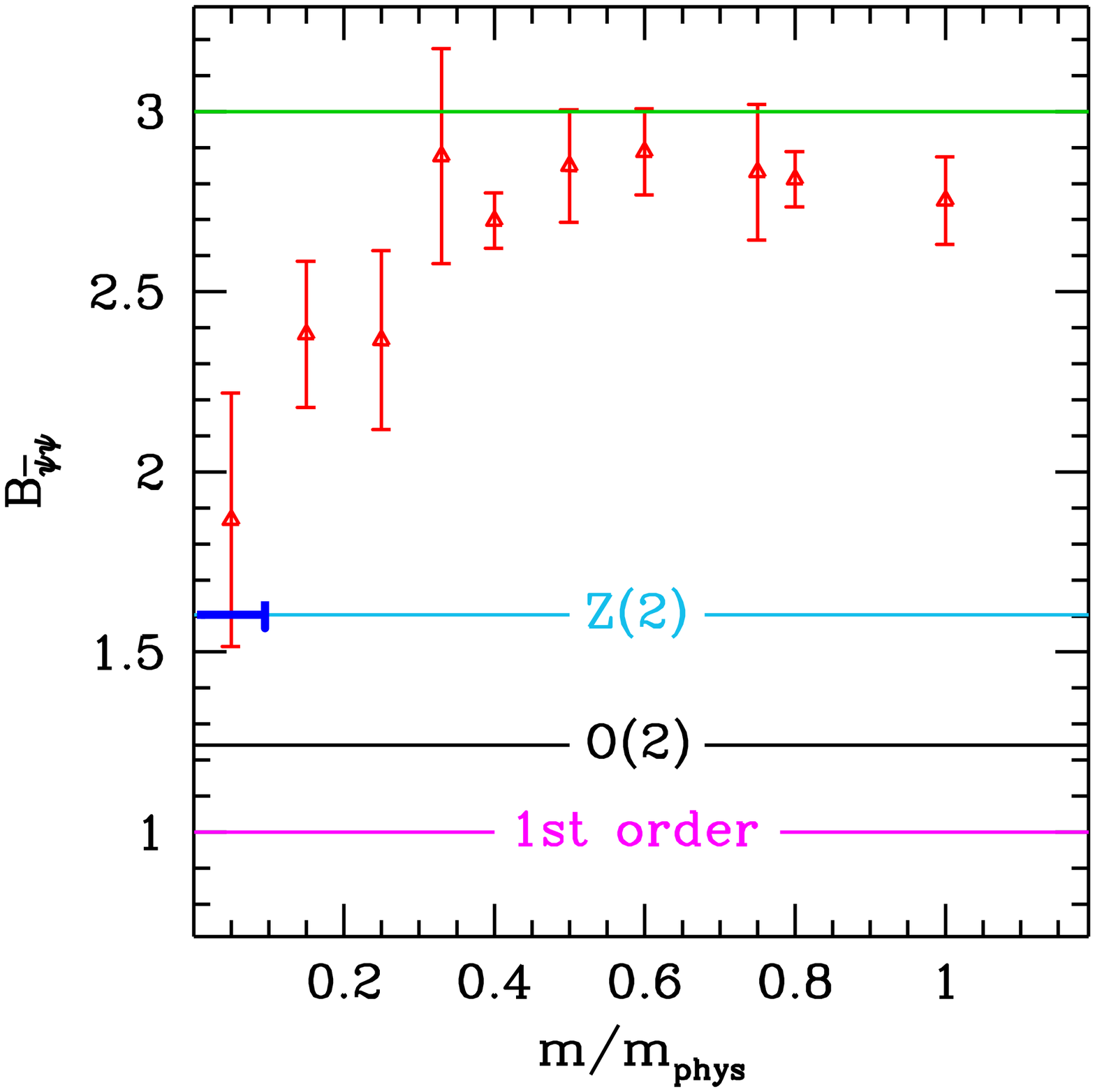}
     	\includegraphics*[height=5cm]{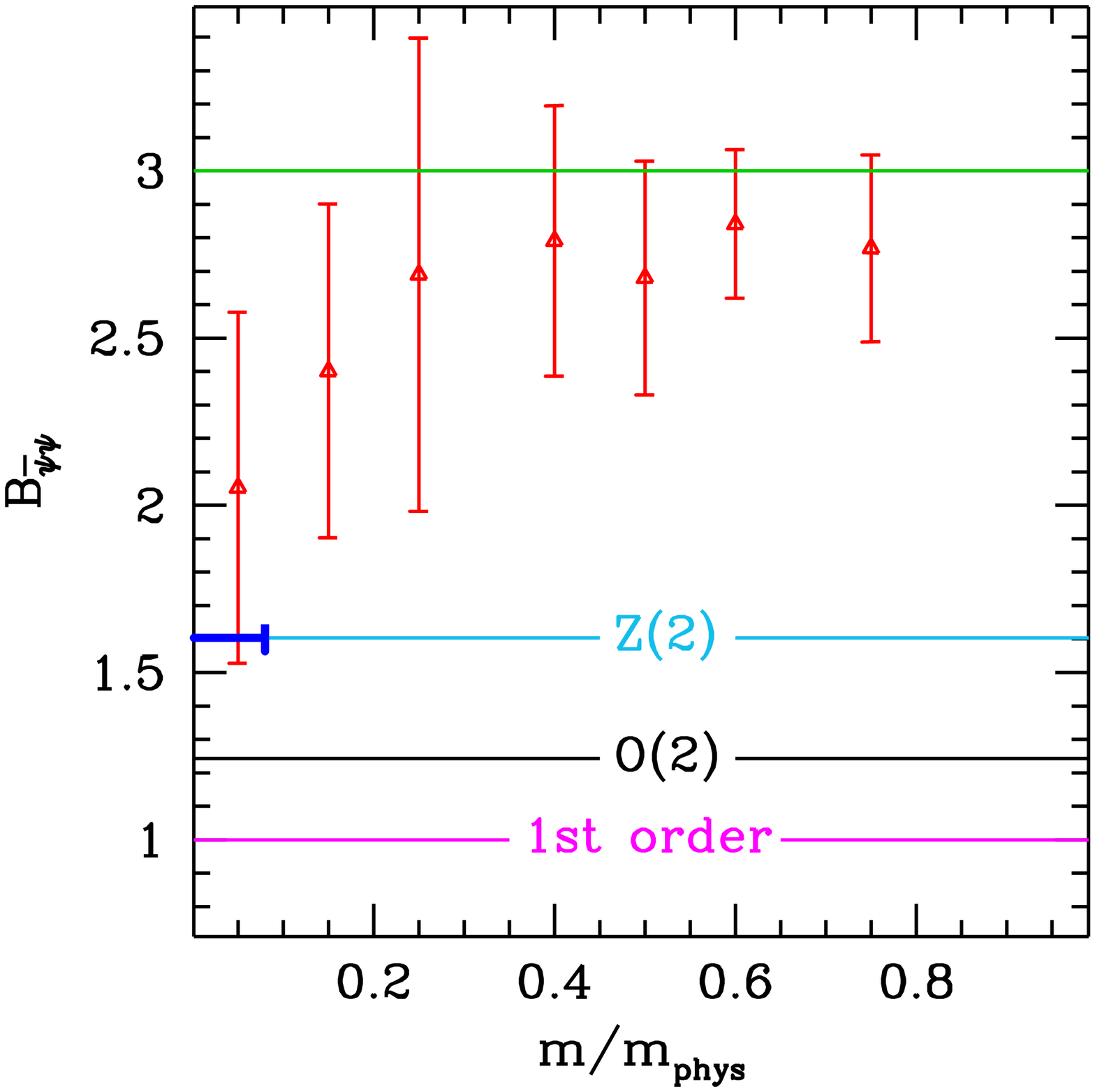}
     	\includegraphics*[height=5cm]{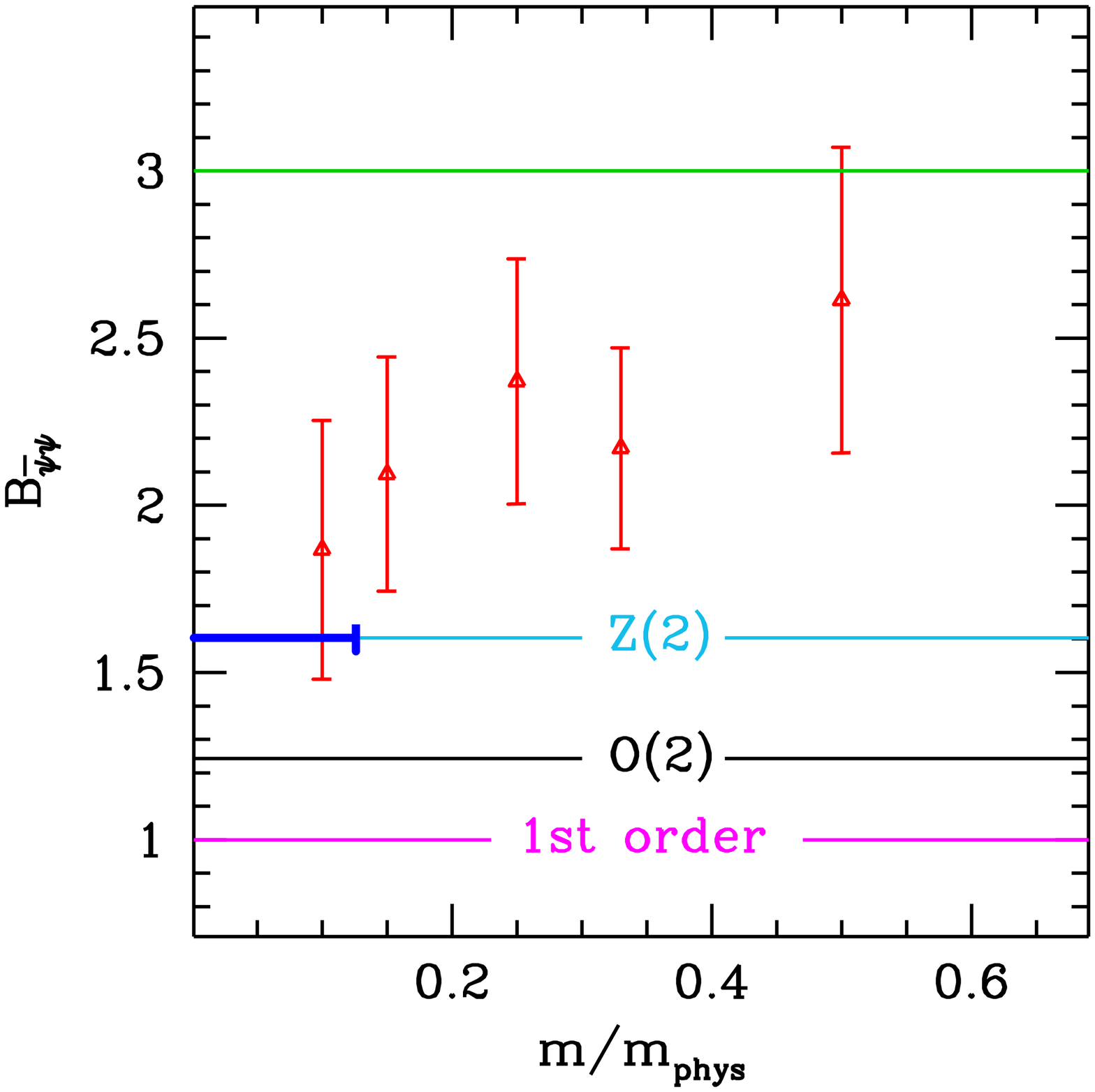}
}
\caption{The value of the Binder cumulant at $T_C$ plotted against the quark mass for $16^3 \times 4$ (left), $24^3 \times 4$ (middle) and $18^3 \times 6$ (right). Behaviour at smaller masses indicate that the universality class of the second-order point in question is consistent with Z(2). We have upper bounds for $m_0$ indicated by the blue lines.}
\label{fig:binder}
\end{figure}

We saw in section \ref{sec:binder}., that the values of the cumulant for the second-order cases and for crossover are quite apart from each other, which makes it easier to have a more accurate estimate for the critical mass. We measured the cumulant at $T_C$, i.e. at $\beta$ corresponding to the position of the peak, which almost always coincided with the minimum of the cumulant in that temperature interval. At larger masses the cumulant has a value consistent with the crossover behaviour, then closer to the critical mass it starts to decrease, and at the point with smallest mass ($m/m_{phys}=0.05$ for $N_T=4$ and $0.1$ for $N_T=6$) it already reaches the value which represents the Z(2) universality class. So the behaviour of the cumulant is consistent with the assumption we posed about figure \ref{fig:quarkmass}. If we accept this scenario, than we can have an upper bound here again for $m_0$. These results are shown on figure \ref{fig:binder}.

From this analysis we can conclude for the critical mass that $m_0/m_{phys}\lesssim 0.07$ for $N_T=4$, and $m_0/m_{phys}\lesssim 0.12$ for $N_T=6$.

\subsection{Summary}

\begin{figure}[h]
\vspace*{-0.2cm}
\centering
\includegraphics[height=5.7cm,width=5.7cm]{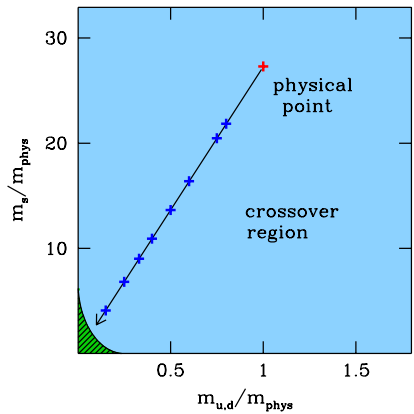}
\caption{The lower left part of the QCD phase diagram, as a conclusion of our work.}
\label{fig:critmass}
\vspace*{-0.8cm}
\end{figure}

The behaviour of the Binder cumulant showed that the universality class of the second-order line is consistent with Z(2). We also obtained upper bounds for the value of the critical mass from the analysis of both the chiral susceptibility and the Binder cumulant. These estimates suggest strongly that the critical mass is below $7 \%$ of the physical quark mass on $N_T=4$ and $12 \%$ on $N_T=6$ lattices. This means that the physical point is at least about ten times farther from the lower left corner of the phase diagram, than the second-order phase line. So the first-order region on figure \ref{fig:quarkmass}. is exaggarated, and looks rather like as de\-pic\-ted on figure \ref{fig:critmass}., which is our final conclusion.

\end{document}